\documentclass[12pt]{article}
\input epsf.tex
\usepackage{cite}
\usepackage{latexsym}
\usepackage{graphicx}
\usepackage{amsmath,amstext,pstricks}
\usepackage{amssymb}
\usepackage{comment}
\numberwithin{equation}{section}
\newcommand{\be}{\begin{equation}}
\newcommand{\ee}{\end{equation}}
\newcommand{\benn}{\begin{equation*}}
\newcommand{\eenn}{\end{equation*}}
\newcommand{\bea}{\begin{eqnarray}}
\newcommand{\eea}{\end{eqnarray}}
\newcommand{\bean}{\begin{eqnarray*}}
\newcommand{\eean}{\end{eqnarray*}}

\def\centeron#1#2{{\setbox0=\hbox{#1}\setbox1=\hbox{#2}\ifdim
\wd1>\wd0\kern.5\wd1\kern-.5\wd0\fi
\copy0\kern-.5\wd0\kern-.5\wd1\copy1\ifdim\wd0>\wd1
\kern.5\wd0\kern-.5\wd1\fi}}
\def\ltap{\;\centeron{\raise.35ex\hbox{$<$}}{\lower.65ex\hbox{$\sim$}}\;}
\def\gtap{\;\centeron{\raise.35ex\hbox{$>$}}{\lower.65ex\hbox{$\sim$}}\;}

\def\lsim{\mathrel{\ltap}}

\begin{document}
\begin{titlepage}
\begin{center}
\hfill RUNHETC-2009-09, SCIPP-09/07 \\

\vskip 0.2in

{\Large \bf Tunneling Constraints on Effective Theories of Stable de Sitter Space}

\vskip 0.3in

Tom Banks$^1$,

\vskip 0.2in

\emph{$^1$ Santa Cruz Institute for Particle Physics,\\
     Santa Cruz, CA 95064\\
     and\\
     Rutgers University NHETC,\\
     Piscataway, NJ 08854}
\vskip0.3in

Jean-Fran\c{c}ois Fortin$^2$,

\vskip 0.2in

\emph{$^2$ Rutgers University NHETC,\\
     Piscataway, NJ 08854}

\begin{abstract}
We argue that effective field theories compatible with the idea of Cosmological SUSY Breaking (CSB), can have {\it no} supersymmetric vacuum states in the $M_P\rightarrow\infty$ limit.  We introduce a revised version of the Pyramid Scheme, which satisfies this criterion.  Combining the criteria for CSB with results of Nelson and Seiberg, any such Lagrangian is non-generic, but we argue that this is plausible in the context of CSB, where R-violating terms in the Lagrangian come from interactions with the horizon, rather than integrating out short distance degrees of freedom.  We also point out a Landau pole in the hidden sector gauge group of the Pyramid Scheme, and propose an unique mechanism for avoiding it.
\end{abstract}

\end{center}
\end{titlepage}

%%%%%%%%%%%%%%%%%%%%%%%%%%%%%%%%%%%%%%%%%%%%%%%%%%%%%%%%%%%%%%%%%%%%%%%%%%%%%%%%%%%%%%%%%%%%%%%%%%%%%%%%%%%%%%%%%%%%%%%%%%%%%%%%%%%%%%%%%%%%%%%%%%%%
%%%%%%%%%%%%%%%%%%%%%%%%%%%%%%%%%%%%%%%%%%%%%%%%%%%%%%%%%%%%%%%%%%%%%%%%%%%%%%%%%%%%%%%%%%%%%%%%%%%%%%%%%%%%%%%%%%%%%%%%%%%%%%%%%%%%%%%%%%%%%%%%%%%%

\section{Introduction}

The central idea of Cosmological SUSY Breaking (CSB) is that the correct quantum theory of stable de Sitter space has the effective cosmological constant (c.c.) as a free parameter.  Supersymmetry is an emergent property of the limit $\Lambda\rightarrow0$, with a scaling relation $m_{3/2}=K  \Lambda^{1/4}$.  Here $K$ is constant of order $10$ \cite{tbaz}.

The basic framework puts strong constraints on the Low Energy Effective Field Theory (LEFT) in the $\Lambda = 0$ limit.  It must be a theory with minimal four dimensional SUSY, as well as an R-symmetry group larger than $Z_2$.  The low energy physics of the model is determined by adding certain R-violating terms to the $\Lambda=0$ Lagrangian, which must give rise to a stable or meta-stable, SUSY violating state, with gravitino mass satisfying the relation above.  Among these terms is a constant superpotential $W_0$, whose function is to tune the effective c.c. to the value indicated by the formula for $m_{3/2}$.  Generally there will also be a SUSY preserving solution of the effective action, with negative cosmological constant.

In gravitational effective field theory, this supersymmetric AdS solution, \textit{has nothing to do with the dS solution}.  It is not part of the same quantum system, which has the de Sitter solution.  This is seen in two complementary ways.  If we consider excitations of the AdS solution, which are normalized and correspond to states, then there are no dS states that are acceptable.  Depending on the scales in the potential, one may create localized excitations with fields concentrated near the positive energy minimum, but as one pushes the size of the region to the dS horizon scale, the excitation becomes a black hole \cite{gftb}.

Correspondingly, if following Coleman and De Luccia (CDL) \cite{cdl}, we look at tunneling from dS space ``to the negative c.c. region", we do not relax to the AdS background, but instead encounter a big crunch, on a microscopic time scale.  Moreover, due to the crunch, the field does not in fact stay localized near the negative c.c. minimum, but instead explores the entire potential until the energy density approaches the Planck scale and effective field theory breaks down.  It seems clear that the non-gravitational effective theory, in which the two solutions correspond to two states of the same Hamiltonian system, with one decaying into the other, is not a correct qualitative description of the physics, even when all the scales are far below the Planck scale.  However, we shall see that the Euclidean solutions of the non-gravitational field theory \textit{are} good approximations to the gravitational CDL instantons, when the range of field variation is small compared to the Planck scale.

As a consequence, {\it we will show that the idea, introduced in \cite{remodel}, of using a meta-stable flat space field theory vacuum state as the LEFT of the theory of stable dS space, is wrong}.  We argue that the only consistent models must use a LEFT which has {\it no} SUSic vacuum in the $M_P\rightarrow\infty$ limit.  Combining the classic results of Nelson and Seiberg \cite{ns} with the basic constraints of CSB, we conclude that the low energy Lagrangian must be non-generic -- that is, it does not include all terms consistent with symmetries, with coefficients determined by dimensional analysis.  We argue that in the context of CSB, the terms in the Lagrangian that violate the fundamental discrete R-symmetry of the $\Lambda=0$ limit, might well be non-generic.  Indeed we already know that this is the explanation, in this context, of the fine tuning of the c.c..  The fundamental requirement for the LEFT is that it reproduces the properties of the underlying quantum theory of stable dS space, which has a finite number of states.  Any CDL instantons must be interpretable as a description of recurrences of low entropy states, rather than true instabilities.

We introduce a modified version of the Pyramid Scheme, with non-generic R-violating terms\footnote{We note that {\it all} versions of the Pentagon model and the Pyramid Scheme secretly invoked the fact that R-violating terms were non-generic, in order to explain the absence of proton decay.}, which has no SUSic vacua.  This model seems to satisfy all the theoretical constraints of CSB and coupling unification, as well as all phenomenological constraints.

We have also taken the opportunity of this paper to repair another flaw that we discovered in the Pyramid Scheme, namely that the hidden sector gauge coupling has a Landau pole below the GUT scale.  The unique way we have discovered to circumvent this is to replace the group at the apex of the Pyramid by $SU_P(4)$, Higgsed to $SU_P(3)$ at about $\lsim50$ TeV.  We have not yet investigated the dynamical source of this Higgs mechanism, but it is perhaps encouraging that it occurs at a scale close to the other scales in the model.  This revision forces us to change the underlying R-symmetry group and the R-charges of various fields.  The simplest model we have found, has a $Z_{13}$ R-symmetry.

%%%%%%%%%%%%%%%%%%%%%%%%%%%%%%%%%%%%%%%%%%%%%%%%%%%%%%%%%%%%%%%%%%%%%%%%%%%%%%%%%%%%%%%%%%%%%%%%%%%%%%%%%%%%%%%%%%%%%%%%%%%%%%%%%%%%%%%%%%%%%%%%%%%%

\subsection{CSB and CDL}

The authors of \cite{abj} showed that the space of potential energy functions for scalar fields, with $|V|>0$ at every minimum, as well as at infinity, could be divided into two classes.  Consider the lowest dS minimum and add a negative constant to the potential to bring this minimum to zero.  The resulting Minkowski solution may or may not have a positive energy theorem, and this is the criterion dividing the two classes.  The co-dimension one dividing line is called The Great Divide.  It is the subspace of potentials that have a static domain wall solution connecting the Minkowski minimum to an AdS solution.  For potentials above the Great Divide, whose Minkowski limit has a positive energy theorem, the probability for the dS ``decay" is of order $e^{- S_{dS}}$, where $S_{dS}$ is the entropy of the de Sitter space.

This is consistent with a model of dS space as a quantum system with a finite number of states \cite{tbwf}, with the ``decay" interpreted as a Poincar\'e recurrence.  The dS vacuum (a high entropy density matrix, not a unique quantum state) is the maximal entropy state of the system, in which the system spends most of its time.  It is properly viewed as stable, despite the existence of the instanton.

This interpretation is consistent with another feature of the instanton solution : the maximal causal diamond in the crunching region of the Lorentzian continuation of the instanton, has an area much smaller than that of the dS horizon.  That is, if we take the holographic interpretation of physics seriously, the instanton is describing a transition from high to low entropy.

By contrast, when the limiting Minkowski vacuum has no positive energy theorem, no such interpretation is possible.  The instanton action is much smaller than the dS entropy and approaches a finite limit as the dS radius goes to infinity.  Thus, the low energy effective theory of a model representing a stable, finite dS universe, must have a potential that is above the Great Divide.

In recent work \cite{pentagon,pyramid}, one of us (T.B.) has been pursuing models of low energy SUSY breaking, which employ the meta-stable states of SUSY-QCD discovered by Intriligator, Seiberg, and Shih \cite{ISS}, and hypothetical generalizations of these states to the theory with an equal number of flavors and colors.  These models have an R-breaking parameter that controls the scale of SUSY breaking.  In the CSB context, one wants to choose this parameter (and the constant in the superpotential) in order to enforce the CSB relation
\begin{equation*}
m_{3/2}=K\Lambda^{1/4},
\end{equation*}
between the gravitino mass and the c.c..  One of us argued that these models were above the Great Divide, because when one dials the R-breaking terms to zero, SUSY is restored and the meta-stable vacuum becomes exactly stable.  \textit{This argument is wrong}.  In this paper, following closely the logic of \cite{abj}, we show that all models in which the potential connecting a meta-stable state to a negative c.c. point, varies rapidly on the Planck scale, are below the Great Divide, and cannot be the low energy effective theory of a stable dS space.

This means that, at the level of non-gravitational effective field theory, the only models compatible with the constraints of CSB, are those which have \textit{no} supersymmetric vacuum states.  Nelson and Seiberg \cite{ns} showed that generic Landau-Ginzburg models of chiral superfields had SUSY preserving minima unless they had an exact $U(1)$ R-symmetry.  Since the rules of CSB require us to break R-symmetry in the LEFT, a generic model can not obey the requirements of CSB.  This may not be as bad as it sounds.  The LEFT of stable dS space has two kinds of terms in its Lagrangian.  The first are terms that exist even in the $\Lambda\rightarrow0$ limit.  These arise through conventional mechanisms and can plausibly be expected to satisfy the requirements of genericity.  On the other hand, there are terms whose sole purpose is to make sure that the physics of the LEFT is compatible with that of the underlying, non-field theoretic, quantum theory of dS space.  At the most fundamental level, it must be compatible with the idea that this system has a finite number of states, the overwhelming majority of which, resemble the dS vacuum.  Transitions out of the dS vacuum should be viewed as recurrences of low entropy states.

In the language of \cite{abj} this means that \textit{LEFTs compatible with CSB must be above the Great Divide}.  Nelson and Seiberg tell us that they must therefore be non-generic.  In the last section of this paper, we will present a modified version of the Pyramid Scheme, with non-generic perturbations, which \textit{is} compatible with CSB.

%%%%%%%%%%%%%%%%%%%%%%%%%%%%%%%%%%%%%%%%%%%%%%%%%%%%%%%%%%%%%%%%%%%%%%%%%%%%%%%%%%%%%%%%%%%%%%%%%%%%%%%%%%%%%%%%%%%%%%%%%%%%%%%%%%%%%%%%%%%%%%%%%%%%
%%%%%%%%%%%%%%%%%%%%%%%%%%%%%%%%%%%%%%%%%%%%%%%%%%%%%%%%%%%%%%%%%%%%%%%%%%%%%%%%%%%%%%%%%%%%%%%%%%%%%%%%%%%%%%%%%%%%%%%%%%%%%%%%%%%%%%%%%%%%%%%%%%%%

\section{Tunneling for meta-stable field theory states}

Consider a model of supersymmetric quantum field theory, with a meta-stable SUSY violating state.  In terms of a (perhaps composite) set of chiral superfields $\{X_i\}$, the superpotential takes the form
\begin{equation*}
W=\mu^3w(X_i/M)+W_0,
\end{equation*}
and the K\"{a}hler potential is
\begin{equation*}
K=M^2k(X_i/M,X_i^*/M).
\end{equation*}
We assume $M\ll m_P$ and $\mu\ll m_P$.  The potential for scalar fields, in SUGRA is then given approximately by
\begin{equation*}
V=\left[\frac{\mu^6}{M^2}|w_i(x^i)|^2-3\frac{|W_0|^2}{m_P^2}\right]\hspace{0.5cm}\mbox{with}\hspace{0.5cm}x^i=\frac{X_i}{M}\hspace{0.5cm}\mbox{and}\hspace{0.5cm}w_i\equiv\frac{\partial w}{\partial x^i}.
\end{equation*}
Naively, this could be a LEFT for CSB if $m_{3/2}=\frac{\mu^3}{M m_P}=K\Lambda^{1/4}$, and $W_0=m_{3/2}m_P^2-\mathcal{O}(\Lambda^{3/4})$.  However, the world is a tough place, and na\"{i}vet\'e often meets with disappointment.  In fact, such a field theory cannot describe the behavior of local excitations of a stable dS space.  To see this, note that the potential has the form $m^4v(X/M)$, where $m=\mu(\mu/M)^{1/2}$.  The tuning of the c.c. implies that the whole potential has this order of magnitude, except right near the meta-stable minimum.

Define $x\equiv X/M$ and re-scale the space time coordinates by the natural time scale $M/m^2$, then the Coleman-DeLucia equations for gravitational tunneling read ($u=-v$) \cite{abj}
\begin{equation*}
\begin{array}{l}
\ddot{x}+3H\dot{x}+u^{\prime}(x)=0\vspace{0.2cm}\\
H=\frac{\dot{r}}{r}\vspace{0.2cm}\\
\dot{r}^2=1+\epsilon^2r^2E\vspace{0.2cm}\\
E=\frac{1}{2}\dot{x}^2+u
\end{array}
\end{equation*}
where $\epsilon^2=M^2/3m_P^2$.  The Euclidean space-time metric is
\begin{equation*}
ds^2=dz^2+\rho^2(z)d\Omega_3^2,
\end{equation*}
and $r$ and the dimensionless Euclidean time $t$ are related to $\rho$ and $z$ by scaling out $\frac{M}{m^2}$.  For the decay of dS space the instanton geometry is an ovoid.  In \cite{abj} it was argued that the situation of a potential w.r.t. the great divide was determined by the stability of the Minkowski solution which is produced when we shift the dS minimum to zero.  In terms of the parameters above, this corresponds to dropping the term of order $\Lambda^{3/4}$ in $W_0$.

If, for the Minkowski limit, we set $\epsilon=0$, then the geometry becomes a semi-infinite cigar.  Coleman \cite{dfv} showed that these equations always have a solution, as long as there is a difference in vacuum energies between the true and false minima.  The asymptotic solution of the scalar field equations approaches the Minkowski stationary point of the potential exponentially fast, which indicates that for very small $\epsilon$ the Minkowski decay occurs in curved dynamical space-time if the corresponding field theoretic decay occurs in Minkowski space.

In \cite{abj}, we showed that for small $\epsilon$ and small positive vacuum energy, one could match this flat space solution to the solution of the field equations in dS space\footnote{See also \cite{bfl}.}.  The instanton manifold is almost the full dS sphere.  As a consequence, the difference between the instanton action and the dS action approaches the flat space instanton action as the dS radius goes to infinity, up to corrections of order $\epsilon^2$.  This shows that if $\epsilon\ll1$ the potential corresponding to a meta-stable vacuum of a non-gravitational field theory is \textit{below the Great Divide}.  Such a potential cannot represent an approximate description of a stable quantum model of dS space.  Indeed, such model has a finite number of states \cite{qds}, the overwhelming majority of which always resemble the dS vacuum.  A small number of states, of order $e^{c(RM_P)^{3/2}}$ represent meta-stable local excitations of the dS vacuum.  CDL decays of such a system, correspond to recurrences of states whose entropy is constant in the limit $RM_P\rightarrow\infty$.  The potential representing such decays must be above the Great Divide.

%%%%%%%%%%%%%%%%%%%%%%%%%%%%%%%%%%%%%%%%%%%%%%%%%%%%%%%%%%%%%%%%%%%%%%%%%%%%%%%%%%%%%%%%%%%%%%%%%%%%%%%%%%%%%%%%%%%%%%%%%%%%%%%%%%%%%%%%%%%%%%%%%%%%
%%%%%%%%%%%%%%%%%%%%%%%%%%%%%%%%%%%%%%%%%%%%%%%%%%%%%%%%%%%%%%%%%%%%%%%%%%%%%%%%%%%%%%%%%%%%%%%%%%%%%%%%%%%%%%%%%%%%%%%%%%%%%%%%%%%%%%%%%%%%%%%%%%%%

\section{Low energy models compatible with CSB}

We are in the fortunate situation of being presented with a paradox.  On the one hand CSB requires the c.c. to be a tunable parameter, which arises at a deeper level as a cosmological initial condition.  For small values of the c.c. the local physics of quantum dS space must be describable in terms of an effective SUGRA Lagrangian with spontaneous SUSY breaking.  The scale of SUSY breaking is $K\Lambda^{1/4}m_P $.  Once we put in the phenomenological lower bounds on superparticle masses, this implies that the mechanism for spontaneous breaking must be understandable in flat space effective field theory.  High scale SUSY breaking by F-terms of moduli fields is not allowed.

The tunneling constraint we have just described implies that the flat space EFT cannot have a SUSic vacuum state, since if it did, it would be below the Great Divide\footnote{One possible loophole in this argument is that a model below the Great Divide, could represent CSB, if the flat space action $\sim(\frac{M}{\mu})^4$ were close to $\pi(RM_P)^2$.  However, since $M\ll M_P$, this can only occur if $\mu\ll\Lambda^{1/4}$, which is inconsistent with experimental lower bounds on super-particle masses.}.  Nelson and Seiberg \cite{ns} have shown that generic chiral Landau-Ginzburg models have SUSic ground states unless the LEFT has a continuous $U_R(1)$.  However, in CSB it is precisely the explicit breaking of R-symmetries that is supposed to trigger SUSY breaking.

In models implementing CSB, the R-axion might also be light enough to cause phenomenological problems, though this depends on the details of the model and assumptions about the scale and dimension of the lowest dimension operator breaking $U_R(1)$.  The universal gravitational contribution, coming from the cancelation of the cosmological constant \cite{bpr}, is too small, given the scale of $W_0$ required in CSB.

One is thus pushed in the direction of assuming a non-generic LEFT.  CSB in fact provides a motivation for non-generic corrections.  Our usual intuition about parameters in effective field theory comes from integrating out high frequency degrees of freedom with the renormalization group.  In CSB, the LEFT has two kinds of terms.  Those that exist in the $\Lambda=0$ limit arise from a model akin to string theory in asymptotically flat space.  They satisfy the usual constraints of effective field theory : generic parameters of order one in appropriate units, consistent with all symmetries.  All mass scales far below the unification scale should be explained dynamically.  By contrast, terms which exist only because of the dS horizon do not obey these rules.  We do not understand the quantum theory of dS space well enough to give a full list of the rules they \textit{do} obey.  We know that the c.c. should be viewed as an input parameter, which means tuning $W_0$ in a way that would be anathema to an effective field theorist.  We know that the new terms should violate R-symmetry, and that their coefficients should enforce the relation $m_{3/2}=K\Lambda^{1/4}$, with $K$ of order $10$.  We have just learned that they must spontaneously break SUSY in a stable vacuum.

Previous work has explored the additional constraints of unification and other aspects of phenomenology.  The constraint that there must be complete multiplets of a GUT group, at the low scale consistent with CSB, and that these new multiplets do not lead to Landau poles in standard model coupling below the unification scale, is very strong and rules out essentially all extant models of gauge mediated or direct mediated SUSY breaking, including the Pentagon model.  These constraints would allow hidden sector gauge groups smaller than $SU_P(5)$, but with a flavor group containing the GUT $SU(5)$ or any larger GUT group, we have not been able to find a model with acceptable dynamics.

The Pyramid Scheme solves this problem by using trinification \cite{trinify}.  GUT multiplets consistent with trinification can add just $D_R$ new vector-like quark multiplets to the colored particle spectrum, where $D_R$ is the representation of the hidden sector gauge group.  In the Pyramid Scheme we chose that group to be $SU(3)$ and $R$ to be the fundamental plus anti-fundamental.  We will see below that this might need to be modified at higher energy.  In the next section we will present a simple generalization of the Pyramid Scheme which satisfies \textit{all} these constraints.

%%%%%%%%%%%%%%%%%%%%%%%%%%%%%%%%%%%%%%%%%%%%%%%%%%%%%%%%%%%%%%%%%%%%%%%%%%%%%%%%%%%%%%%%%%%%%%%%%%%%%%%%%%%%%%%%%%%%%%%%%%%%%%%%%%%%%%%%%%%%%%%%%%%%
%%%%%%%%%%%%%%%%%%%%%%%%%%%%%%%%%%%%%%%%%%%%%%%%%%%%%%%%%%%%%%%%%%%%%%%%%%%%%%%%%%%%%%%%%%%%%%%%%%%%%%%%%%%%%%%%%%%%%%%%%%%%%%%%%%%%%%%%%%%%%%%%%%%%

\section{Pyramid Schemes with a triplet of singlets}

The new chiral matter content of the Pyramid Scheme consists of a singlet $S$ and three chiral pairs ${\cal T}_i, \tilde{\cal T}_i$.  The gauge group is $SU(3)^4\rtimes Z_3$.  The first $SU(3)$, called $SU_P(3)$, is the hidden sector gauge group, while the rest forms Glashow's trinification group, in which the $Z_3$ permutes the three $SU(3)$ factors, ensuring coupling unification at the GUT scale.  We will be working at energies far below the GUT scale, where this group is broken to the $SU(1,2,3)$ of the standard model.  We label the three $SU(3)$ groups of trinification $SU_i(3)$, with $i=1,2,3$.  For $i=2,3$, the $SU(i)$ of the standard model is the obvious Cartesian subgroup of $SU_i(3)$.  Weak hypercharge is a linear combination of a generator of $SU_1(3)$ with the hypercharge generator in $SU_2(3)$.  We will occasionally write terms in the Lagrangian that preserve more of the GUT symmetry than is required by general principles.  We do this for convenience only.  We believe that, as long as we do not introduce huge differences between parameters that are set equal by this choice, the qualitative physics of our model will remain unchanged.  Another way to say this is that we have found a variety of Pyramid Schemes, with multiple parameters, which satisfy all of our fundamental and phenomenological constraints.  For economy's sake we only write down the simplest one explicitly.

The new models we introduce in this paper replace the singlet $S$ by a triplet of singlets $S_i$ with $i=1,\ldots,3$.  We imagine that, neglecting GUT symmetry breaking, these triplets transform into each other under the $Z_3$.  However, in this paper we will not attempt to write down a GUT field theory or string compactification which reduces to our model below the scale of GUT symmetry breaking.  When the c.c. $\Lambda = 0$, the $S_i$ appear in the superpotential as
\begin{equation*}
W_{\{S_i\}}=y_iS_i{\cal T}_i\tilde{\cal T}_i+\beta_iS_iH_u H_d,
\end{equation*}
with repeated indices summed.

When $\Lambda$ is turned on, we add the terms
\begin{equation*}
m_i{\cal T}_i\tilde{\cal T}_i+M^2_iS_i.
\end{equation*}
The coefficients in these terms will scale to zero with $\Lambda$ and are chosen to enforce the relation $m_{3/2}=K\Lambda^{1/4}$.

At high energies, the hidden sector is SUSY QCD with $9$ flavors and $3$ colors\footnote{In order to avoid Landau poles in the hidden sector gauge coupling, we will later contemplate an enhanced hidden sector gauge symmetry, reduced to this one by the Higgs mechanism at a fairly high scale.}.  This model has a vanishing one loop beta function, which is positive at two loops.  Thus the coupling slowly decreases as we go down in energy scale.  We will assume that $m_{1,3}$ are both $>m_2$.  After integrating out the heavy trianons, we have the $N_F=N_C=3$ model, and we assume that this becomes strongly coupled at a scale $\Lambda_3$ just below $m_2$.

Now let us discuss candidates for the discrete R-symmetry which is part of the rules of the game of CSB.  The $(3,9)$ gauge theory has an anomaly free $U_R(1)$ symmetry under which all the trianon and anti-trianon fields have charge $2/3$.  We can choose a discrete subgroup of this, and add any cyclic subgroup of the $SU_L(9)\times SU_R(9)\times U_B(1)$ flavor group.  We must check that the symmetry is not broken by standard model instantons.  Finally, we want to reproduce the success of previous models and use this symmetry to forbid all dimension four and five operators in the MSSM, which violate $B$ or $L$, apart from the neutrino seesaw operators $(LH_u)^2$.  We know that this can be accomplished if we choose ${\cal T}_2$, $\tilde{\cal T}_2$ to have R-charge $0$ and $S_2$ to have R-charge $2$.

The low energy superpotential is written in terms of the fields $S_i$, $H_u$, $H_d$ and the mesons and baryons of the gauge theory.  We parametrize the dimension one meson matrix by
\begin{equation*}
M=Ze^{\frac{\lambda_aZ_a}{\Lambda_3}},
\end{equation*}
where $\lambda_a$ are the eight traceless Gell-Mann matrices.  We will search for $SU(3)$ symmetric stationary points, where $Z_a=0$.  The superpotential is
\begin{equation*}
W=3\Lambda_3(m_2+y_2S_2)Z+L(Z^3/\Lambda_3-B\tilde{B}-\Lambda_3^2)+\beta_iS_iH_uH_d+M_i^2S_i.
\end{equation*}
The equations from the variation of $B$ and $\tilde{B}$ either force these fields to be zero or $L$ to vanish.  We explore the second alternative first.  The variation of $Z$, for $L=0$ implies
\begin{equation*}
y_2S_2+m_2=0.
\end{equation*}
The variational equations for the $S_i$ imply
\begin{equation*}
3\delta_{i2}y_2\Lambda_3Z+\beta_iH_uH_d+M_i^2=0.
\end{equation*}
These are three equations for two unknowns, and have no solution.

Turning to the solution $B=\tilde{B}=0$, we note that the moduli space constraint now freezes $Z^3=\Lambda_3^3$, which has $3$ solutions.  The $Z$ equation fixes $L$ in terms of $S_2$, but the $S_i$ equations are now three equations for the single unknown $H_uH_d$.  Therefore, we do not find any supersymmetric solution on either branch of the moduli space.

This conclusion is unchanged if we explore non-zero values of the adjoint fields $Z_a$.  These appear only through a multiplicative factor ${\rm Tr}\ e^{\lambda_aZ_a}$ in the term in the superpotential linear in $Z$.  The variational equations for these fields are of course satisfied when $Z_a=0$, and there are other solutions.  If we are on the branch where $L=0$ then all values of the $Z_a$ are stationary.  None of this changes the fact that there are no solutions of the variational equations for the $S_i$.

We note that it is the parameters $M_i^2$ which prevent us from having a supersymmetric solution.  If they vanished, then on the branch with $L=0$ we can solve the $S_2$ equation by fixing $y_2\Lambda_3Z+\beta_2H_uH_d=0$, and the other equations are both solved by $H_u=H_d=0$ (which also solves the variational equations for the Higgs fields -- equations we have not yet discussed).  Therefore, the crucial SUSY violating equations are those which come from varying $S_{1,3}$.

%%%%%%%%%%%%%%%%%%%%%%%%%%%%%%%%%%%%%%%%%%%%%%%%%%%%%%%%%%%%%%%%%%%%%%%%%%%%%%%%%%%%%%%%%%%%%%%%%%%%%%%%%%%%%%%%%%%%%%%%%%%%%%%%%%%%%%%%%%%%%%%%%%%%
%%%%%%%%%%%%%%%%%%%%%%%%%%%%%%%%%%%%%%%%%%%%%%%%%%%%%%%%%%%%%%%%%%%%%%%%%%%%%%%%%%%%%%%%%%%%%%%%%%%%%%%%%%%%%%%%%%%%%%%%%%%%%%%%%%%%%%%%%%%%%%%%%%%%

\section{$SU_P(3)$ Landau pole and $SU_P(4)$ completion}

It is interesting to look at the strongly-coupled gauge theory beta function since this sector consists of $SU_P(N_C)$ SQCD with $N_C=3$ and $N_F=9$ and is thus not asymptotically free.  One can therefore ask what are the lightest ISS masses compatible with a strongly-coupled $SU_P(3)$, such that there is no Landau pole below the GUT scale.  As we mentioned earlier, the resulting large ISS mass hierarchy will suggest that we look instead at $SU_P(4)$ which is Higgsed to $SU_P(3)$ at some high scale.  To perform the analysis, it is convenient to look at the general case of $SU_P(N_C)$ SQCD with $N_F$ flavors.  The $\beta$-function for $SU(N_C)$ with $N_F$ fundamental flavors is
\begin{eqnarray*}
\beta_g &=& -\frac{g^3}{16\pi^2}\frac{3N_C-N_F+N_F\gamma}{1-N_C\frac{g^2}{8\pi^2}}\\
\gamma &=& -\frac{g^2}{8\pi^2}\frac{N_C^2-1}{N_C}+\mathcal{O}(g^4)
\end{eqnarray*}
or in terms of the fine structure constant $\alpha=g^2/4\pi$
\begin{eqnarray*}
\beta_{\frac{2\pi}{\alpha}} &=& \frac{3N_C-N_F+N_F\gamma}{1-N_C\frac{\alpha}{2\pi}}\\
\gamma &=& -\frac{\alpha}{2\pi}\frac{N_C^2-1}{N_C}+\mathcal{O}(\alpha^2)
\end{eqnarray*}
At first order, the solution is
\begin{eqnarray*}
\frac{2\pi}{\alpha(\mu)} &=& \frac{2\pi}{\alpha(\mu_0)}+(3N_C-N_F)\ln(\mu/\mu_0)\hspace{1cm}\mbox{for $N_F\neq3N_C$}\\
\left(\frac{2\pi}{\alpha(\mu)}\right)^2 &=& \left(\frac{2\pi}{\alpha(\mu_0)}\right)^2-6(N_C^2-1)\ln(\mu/\mu_0)\hspace{1cm}\mbox{for $N_F=3N_C$}
\end{eqnarray*}
In our case, we expect the hierarchy $\Lambda_3<m_2<m_3<m_1<M_{{\rm GUT}}$ where $m_3$ cannot be too much larger than $\Lambda_3$ due to the experimental lower bound on the gluino mass.  When $m_3$ is large, there are no light messengers, which carry color.  Thus the strongly-coupled theory has $0$ flavors between $\Lambda_3$ and $m_2$, $3$ flavors between $m_2$ and $m_3$, $6$ flavors between $m_3$ and $m_1$ and $9$ flavors between $m_1$ and $M_{{\rm GUT}}$.  At leading order, this leads to
\begin{eqnarray*}
\left(\frac{2\pi}{\alpha(\mu)}\right)^2 &=& \left[9\ln(m_2/\Lambda_3)+6\ln(m_3/m_2)+3\ln(m_1/m_3)\right]^2-48\ln(\mu/m_1)\\
 &=& 9\ln^2(m_1m_2m_3/\Lambda_3^3)-48\ln(\mu/m_1)
\end{eqnarray*}
where $m_1<\mu<M_{{\rm GUT}}$.  With the generic numbers $\Lambda_3=5$ TeV, $m_2=9$ TeV and $m_3=12$ TeV, asking for the Landau pole to be above the GUT scale leads to $m_1\gtrsim4\times10^4$ TeV.  This is quite a large hierarchy of scales for the ISS masses.  Indeed it is so large that it ruins standard model gauge coupling unification.  With this spectrum of trianons, in the one loop approximation, $\alpha_1(M_{{\rm GUT}})$ is $\sim20$ \% away from the value it should be for unification.

The best way to circumvent this hierarchy is to assume that the theory is an $SU_P(4)$ with $N_F=9$ flavors which is Higgsed to the previous $SU_P(3)$ with $N_F=9$ flavors at a scale determined by the VEV $V_4$ of a chiral field in the $N_F = 9$, $N_C = 4$ model.  Now the (mild) hierarchy of scales becomes $\Lambda_3<m_2<m_3<m_1<V_4<M_{{\rm GUT}}$.  Following the same analysis as shown above, the constraint on the VEV follows from
\begin{equation*}
\frac{2\pi}{\alpha(\mu)}=\sqrt{9\ln^2(m_1m_2m_3/\Lambda_3^3)-48\ln(V_4/m_1)}+3\ln(\mu/V_4)
\end{equation*}
where $V_4<\mu<M_{{\rm GUT}}$.

With $\Lambda_3=5$ TeV, $m_2=9$ TeV, $m_3=12$ TeV and $m_1=15$ TeV the theory is well-behaved for $V_4\lsim50$ TeV.  The VEV cannot be pushed to very high scales due to the behavior of the beta function when $N_F=3N_C$.

There are certainly loci on the moduli space of the $N_F=9$, $N_C=4$ theory with this pattern of Higgs VEVs.  We have not investigated the origin of the potential which might fix the theory at such a point.  We have thus exhibited two possible mechanisms for avoiding a Landau pole in the $SU_P(3)$ coupling below the GUT scale, but only one consistent with standard model gauge coupling unification.  The enhancement of the hidden sector gauge group introduces scales in the same ballpark as the rest of the energy scales in the model.  Adopting it, we incur a debt to explain a new $10 - 100$ TeV scale Higgs mechanism, which we hope to repay at a later date.

%%%%%%%%%%%%%%%%%%%%%%%%%%%%%%%%%%%%%%%%%%%%%%%%%%%%%%%%%%%%%%%%%%%%%%%%%%%%%%%%%%%%%%%%%%%%%%%%%%%%%%%%%%%%%%%%%%%%%%%%%%%%%%%%%%%%%%%%%%%%%%%%%%%%
%%%%%%%%%%%%%%%%%%%%%%%%%%%%%%%%%%%%%%%%%%%%%%%%%%%%%%%%%%%%%%%%%%%%%%%%%%%%%%%%%%%%%%%%%%%%%%%%%%%%%%%%%%%%%%%%%%%%%%%%%%%%%%%%%%%%%%%%%%%%%%%%%%%%

\section{Discrete R-symmetry}

The R-charges in the Pyramid Scheme with a triplet of singlets follow the usual rules.  Here we look for a R-charge assignment which leads to the vanishing of the 't Hooft operators for $SU_P(4)$.  Another constraint comes from the trilinear singlet-Higgs couplings, $S_i H_u H_d$ which cannot be in the Lagrangian for all $i=1,\ldots,3$.  If they were, all singlets $S_i$ would have the same R-charge and this is prohibited by the vanishing of the $SU_C(3)$ 't Hooft operator.  Therefore one has to choose $\beta_3=0$ with $\beta_{i=1,2}$ arbitrary and then the $S_{i=1,2}$ singlets share the same R-charge.  Denoting the R-charge of a field by the field itself, this implies $S_1=S_2\equiv S$.  In the GUT notation, the extra matter fields are
\begin{equation*}
\begin{array}{c|cccc}
 & SU_1(3) & SU_2(3) & SU_3(3) & SU_P(3)\\\hline
{\cal T}_1 & 3 & 1 & 1 & \bar{3}\\
\bar{\cal T}_1 & \bar{3} & 1 & 1 & 3\\
{\cal T}_2 & 1 & 3 & 1 & \bar{3}\\
\bar{\cal T}_2 & 1 & \bar{3} & 1 & 3\\
{\cal T}_3 & 1 & 1 & 3 & \bar{3}\\
\bar{\cal T}_3 & 1 & 1 & \bar{3} & 3\\
S_{i=1,2,3} & 1 & 1 & 1 & 1.
\end{array}
\end{equation*}
As explained above, our goal is to find an approximate discrete R-symmetry which is exact in the limit of zero c.c. and which allows only the terms needed in this limit.  To simplify the search, we will look at a continuous $U_R(1)$, of which we imagine only a discrete $Z_N$ subgroup is fundamental.  Therefore, all the following equations only have to be satisfied modulo $N$.

The only superpotential terms which are required at the renormalizable level are
\begin{equation*}
W_{\Lambda=0}\supset
S_i{\cal T}_i\bar{\cal T}_i,\;S_{i=1,2}H_uH_d,\;H_uQ\bar{U},\;H_dQ\bar{D},\;H_dL\bar{E},\;(LH_u)^2
\end{equation*}
which implies that the R-charges satisfy
\begin{eqnarray*}
{\cal T}_{i=1,2}+\bar{\cal T}_{i=1,2} &=& 2-S\\
{\cal T}_3+\bar{\cal T}_3 &=& 2-S_3\\
H_u &=& 2-H_d-S\\
\bar{U} &=& H_d+S-Q\\
\bar{D} &=& 2-H_d-Q\\
\bar{E} &=& 2-H_d-L
\end{eqnarray*}
since $S_{i=1,2}\equiv S$ and the extra relation from the neutrino seesaw operator has still to be taken into account.  All remaining renormalizable superpotential terms must be forbidden by the discrete R-symmetry otherwise we would expect them to be in the superpotential with order 1 coefficients in the appropriate units.  Moreover, dangerous higher-dimensional $B$ and $L$ violating terms must be forbidden as well by the discrete R-symmetry to insure proton stability on appropriate timescales.

The (approximate) $U_R(1)$ anomaly conditions are
\begin{eqnarray*}
SU_P(4)^2U_R(1) &\Rightarrow& 2\cdot4+3({\cal T}_1+\bar{\cal T}_1+{\cal T}_2+\bar{\cal T}_2+{\cal T}_3+\bar{\cal T}_3-6)\\
 && \hspace{1cm}=8-6S-3S_3\\
SU_C(3)^2U_R(1) &\Rightarrow& 2\cdot3+6(Q-1)+3(\bar{U}+\bar{D}-2)+4({\cal T}_3+\bar{\cal T}_3-2)\\
 && \hspace{1cm}=3S-4S_3\\
SU_L(2)^2U_R(1) &\Rightarrow& 2\cdot2+(H_u+H_d-2)+9(Q-1)+3(L-1)\\
 && \hspace{1cm}+4({\cal T}_2+\bar{\cal T}_2-2)=3(3Q+L)-8-5S.
\end{eqnarray*}
These lead to the equation $S_3=9S-8$ and the 't Hooft constraints
\begin{eqnarray*}
32-33S &=& 0\\
3(3Q+L)-8-5S &=& 0.
\end{eqnarray*}

The dangerous renormalizable and higher-dimensional $B$ and $L$ violating superpotential and K\"{a}hler potential terms (note that the neutrino seesaw operator is required) can be combined into $13$ groups,
\begin{eqnarray*}
G_1=\{LL\bar{E},\;LQ\bar{D},\;SLH_u\} &\Rightarrow& L-H_d\\
G_2=\{\bar{U}\bar{D}\bar{D}\} &\Rightarrow& 3Q+H_d-S-2\\
G_3=\{LH_u,\;Q\bar{U}\bar{E}H_d,\;\bar{U}\bar{D}^*\bar{E},\;H_u^*H_d\bar{E},\;Q\bar{U}L^*\} &\Rightarrow& L-H_d-S\\
G_4=\{S_3LH_u\} &\Rightarrow& L-H_d+8S-8\\
G_5=\{QQQL\} &\Rightarrow& 3Q+L-2\\
G_6=\{QQQH_d,\;QQ\bar{D}^*\} &\Rightarrow& 3Q+H_d-2\\
G_7=\{\bar{U}\bar{U}\bar{D}\bar{E}\} &\Rightarrow& 3Q+L-2S-2\\
G_8=\{LH_uH_dH_u\} &\Rightarrow& L-H_d-2S+2\\
G_9=\{SLL\bar{E},\;SLQ\bar{D},\;S^2LH_u\} &\Rightarrow& L-H_d+S\\
G_{10}=\{S\bar{U}\bar{D}\bar{D}\} &\Rightarrow& 3Q+H_d-2S-2\\
G_{11}=\{SS_3LH_u,\;S_3LL\bar{E},\;S_3LQ\bar{D}\} &\Rightarrow& L-H_d+9S-8\\
G_{12}=\{S_3\bar{U}\bar{D}\bar{D}\} &\Rightarrow& 3Q+H_d-10S+6\\
G_{13}=\{S_3^2LH_u\} &\Rightarrow& L-H_d+17S-16.
\end{eqnarray*}

Moreover, the forbidden renormalizable superpotential terms can be combined into $12$ groups,
\begin{eqnarray*}
G_{14}=\{{\cal T}_1\bar{\cal T}_1,\;{\cal T}_2\bar{\cal T}_2,\;H_uH_d\} &\Rightarrow& S\\
G_{15}=\{{\cal T}_3\bar{\cal T}_3\} &\Rightarrow& 9S-8\\
G_{16}=\{S_3H_uH_d\} &\Rightarrow& 8S-8\\
G_{17}=\{S\} &\Rightarrow& S-2\\
G_{18}=\{S^2\} &\Rightarrow& 2S-2\\
G_{19}=\{S^3\} &\Rightarrow& 3S-2\\
G_{20}=\{S_3\} &\Rightarrow& 9S-10\\
G_{21}=\{S_3^2\} &\Rightarrow& 18S-18\\
G_{22}=\{S_3^3\} &\Rightarrow& 27S-26\\
G_{23}=\{SS_3\} &\Rightarrow& 10S-10\\
G_{24}=\{S^2S_3\} &\Rightarrow& 11S-10\\
G_{25}=\{SS_3^2\} &\Rightarrow& 19S-18.
\end{eqnarray*}

Operators in each group have the same R-charge (once one takes the $d^2\theta$ for superpotential terms into account).  It is possible to forbid all dangerous terms with $N=13$, and $S=12$, $Q=0$, $L=1$, and $H_d=3$.  With this choice all anomaly conditions are satisfied, only the required terms do not break the discrete R-symmetry and thus none of the dangerous terms are allowed.  Notice moreover that the neutrino seesaw operator is allowed as required by this choice of R-charges.  Therefore one can engineer a generic superpotential of the form
\begin{multline*}
W_{\Lambda=0}=\sum_{i=1}^3y_iS_i{\cal T}_i\bar{\cal T}_i+\sum_{i=1}^2\beta_iS_iH_uH_d\\
+\lambda_uH_uQ\bar{U}+\lambda_d H_dQ\bar{D}+\lambda_LH_dL\bar{E}+\frac{\lambda_{\nu}}{m_P}(L H_u)^2
\end{multline*}
which is supplemented by the non-generic superpotential
\begin{equation*}
\delta W_{\Lambda\neq0}=\sum_{i=1}^3\left(m_i{\cal T}_i\bar{\cal T}_i+M_i^2S_i\right)+W_0
\end{equation*}
when the c.c. is turned on.  Note that in these equations $\lambda_{u,d,L,\nu}$ are all matrices in generation space.

%%%%%%%%%%%%%%%%%%%%%%%%%%%%%%%%%%%%%%%%%%%%%%%%%%%%%%%%%%%%%%%%%%%%%%%%%%%%%%%%%%%%%%%%%%%%%%%%%%%%%%%%%%%%%%%%%%%%%%%%%%%%%%%%%%%%%%%%%%%%%%%%%%%%
%%%%%%%%%%%%%%%%%%%%%%%%%%%%%%%%%%%%%%%%%%%%%%%%%%%%%%%%%%%%%%%%%%%%%%%%%%%%%%%%%%%%%%%%%%%%%%%%%%%%%%%%%%%%%%%%%%%%%%%%%%%%%%%%%%%%%%%%%%%%%%%%%%%%

\section{Conclusions}

When combined with fairly broad brush phenomenological requirements, the idea of CSB is constrained in quite a remarkable manner.  The strongest constraints come from the combination of the low scale of SUSY breaking required by CSB, and coupling unification.  Most models of gauge mediation, both direct and with an intermediate messenger sector, are ruled out.  The only models we have found, which satisfy these constraints, are variations on the Pyramid Scheme.

In this paper, we pointed out two new constraints and proposed a class of models that satisfies them.  The first constraint comes from the fundamental requirement that the LEFT of a theory of stable dS space, {\it must} be above the Great Divide.  On the other hand, we showed that flat space field theory models, with a meta-stable SUSY violating vacuum and a SUSic vacuum a distance $M \ll m_P$ away in field space (and no unnatural fine tuning besides the tuning of the c.c.), are all below the Great Divide.

CSB {\it requires} very low scale SUSY breaking, so the only way to achieve this is for the LEFT to have no SUSic vacuum at all.  The seminal paper of Nelson and Seiberg shows us that this is achievable in a generic manner, only if the model has an unbroken $U_R (1)$, which is spontaneously broken.  This however is {\it incompatible} with the requirements of CSB, according to which a discrete R-symmetry and a SUSic vacuum are both restored in the $\Lambda=0$ limit.  Explicit R-violating terms are then supposed to remove the SUSic vacuum.  We argued that there is no reason to assume those R-violating terms obeyed the rules of quantum field theory naturalness.  We exhibited an explicit variation on the Pyramid Scheme, with a separate singlet for each leg of the Pyramid, which satisfied all these requirements.

The second issue we studied was the occurrence of Landau poles below the GUT scale in the hidden sector gauge coupling.  We argued that to avoid these, preserving the phenomenological successes of the model, we either had to take one R-violating trianon mass to be very large $\gtrsim4\times10^4$ TeV, or embed $SU_P(3)$ in $SU_P(4)$ with a Higgs mechanism at $\lsim50$ TeV.  However, the first idea ruins standard model gauge coupling unification.  Thus, the only scheme consistent with CSB, with gauge coupling unification, and with standard model phenomenology is a pyramid with an $SU(4)$ apex, reduced to the $N_F = N_C = 3$ model by a combination of the Higgs mechanism and trianon masses.  All of the scales of the model are in the $1 - 100$ TeV regime.  We have not yet investigated the dynamical mechanism which could account for this new Higgs mechanism, which breaks $SU_P(4)$ to $SU_P(3)$.

Finally, everything is connected in the Pyramid Scheme, and we were forced to revisit the issue of the discrete R-symmetry group and its role in suppressing dimension four and five operators that violate $B$ and $L$. The simplest model we found uses a $Z_{13}$ R-symmetry group.

%%%%%%%%%%%%%%%%%%%%%%%%%%%%%%%%%%%%%%%%%%%%%%%%%%%%%%%%%%%%%%%%%%%%%%%%%%%%%%%%%%%%%%%%%%%%%%%%%%%%%%%%%%%%%%%%%%%%%%%%%%%%%%%%%%%%%%%%%%%%%%%%%%%%
%%%%%%%%%%%%%%%%%%%%%%%%%%%%%%%%%%%%%%%%%%%%%%%%%%%%%%%%%%%%%%%%%%%%%%%%%%%%%%%%%%%%%%%%%%%%%%%%%%%%%%%%%%%%%%%%%%%%%%%%%%%%%%%%%%%%%%%%%%%%%%%%%%%%

\section{Acknowledgments}

We would like to thank K. Intriligator for conversations about SUSY breaking and R-symmetry.  This research was supported in part by DOE grant number DE-FG02-96ER40949 and DOE grant number DE-FG03-92ER40689.

%%%%%%%%%%%%%%%%%%%%%%%%%%%%%%%%%%%%%%%%%%%%%%%%%%%%%%%%%%%%%%%%%%%%%%%%%%%%%%%%%%%%%%%%%%%%%%%%%%%%%%%%%%%%%%%%%%%%%%%%%%%%%%%%%%%%%%%%%%%%%%%%%%%%
%%%%%%%%%%%%%%%%%%%%%%%%%%%%%%%%%%%%%%%%%%%%%%%%%%%%%%%%%%%%%%%%%%%%%%%%%%%%%%%%%%%%%%%%%%%%%%%%%%%%%%%%%%%%%%%%%%%%%%%%%%%%%%%%%%%%%%%%%%%%%%%%%%%%

\end{document}